\documentclass[aps,pra,twocolumn,floatfix,superscriptaddress,showpacs,showdate]{revtex4}

\usepackage{graphicx}
\usepackage{bbm}
\usepackage{amsmath,amssymb}
\usepackage{color,calc}

\newcommand{\be}{\begin{equation}}
\newcommand{\ee}{\end{equation}}
\newcommand{\bea}{\begin{eqnarray}}
\newcommand{\eea}{\end{eqnarray}}

\def\d{\delta}
\def\D{\Delta}

\def\ve{\varepsilon}

\def\w{\omega}

\def\bra{\langle}
\def\ket{\rangle}

\def\xc{{\rm xc}}
\def\x{{\rm x}}

\begin{document}

\title{Optimal control of strong-field ionization with time-dependent \\density-functional theory}

\author{Maria Hellgren}
\affiliation{Max-Planck Institute of
Microstructure Physics, Weinberg 2, 06120 Halle, Germany}
\affiliation{SISSA, International School for Advanced Studies, 
via Bonomea 265, 34136 Trieste, Italy}
\author{Esa R{\"a}s{\"a}nen}
\affiliation{Department of Physics, Tampere University of Technology, 
FI-33101 Tampere, Finland}
\author{E. K. U. Gross}
\affiliation{Max-Planck Institute of
Microstructure Physics, Weinberg 2, 06120 Halle, Germany}
\date{\today}

\pacs{31.15.ee, 31.15.vj, 32.80.Qk, 33.80.Rv}

\begin{abstract}
We show that quantum optimal control theory (OCT) and time-dependent 
density-functional theory (TDDFT) can be combined to provide 
realistic femtosecond laser pulses for an enhanced ionization 
yield in many-electron systems. Using the H$_2$-molecule as a test 
case, the optimized laser pulse from the numerically exact scheme 
is compared to pulses obtained from OCT+TDDFT within the TD 
exact-exchange (TDEXX) and the TD local-density approximation 
(TDLDA). 
We find that the TDDFT-pulses produces an ionization yield of up to $50\%$ 
when applied to the exact system. In comparison, pulses with
a single frequency but the same fluence typically reach to
yields around $5-15\%$, unless the frequency is carefully tuned
into a Fano-type resonance that leads to $\sim 30\%$ yield. 
On the other hand, optimization within the exact system alone leads 
to yields higher than $80\%$, demonstrating that correlation effects 
beyond the TDEXX 
and TDLDA can give rise to even more efficient ionization mechanisms. 
\end{abstract}

\maketitle

\section{Introduction}

Developments in ultrafast science on the electronic timescale have
been impressive during the past few years~\cite{Krausz2009}.
In particular, manipulation of both intense infrared and 
weak extreme ultraviolet pulses has led to innovative schemes to measure
time delays in the attosecond range~\cite{Schultze2010,Klunder2011}.
It is expected that the ongoing advances will soon open a path
into monitoring and controlling real-time electron dynamics.

In addition to measurements on the electronic timescale, subcycle
pulse shaping has recently become possible. Wirth and 
co-workers~\cite{Wirth2011} have generated synthesized laser pulses
by combining subcycle transients across the infrared, visible, and
ultraviolet regimes. In each regime, respectively, the chirp, carrier
envelope phase, time delay, and energy (beam size) can be
controlled before the final pulse is reconstructed.

The advances mentioned above bring the applications of optimized
control schemes -- such as quantum optimal control 
theory~\cite{oct,octreview1,octreview2} (OCT) -- to a new level of practical
relevance. In atomic physics, OCT has been previously applied to design
laser pulses for, e.g., enhanced~\cite{alberto} and suppressed 
ionization~\cite{lars}. For molecular processes the range of applications
is significantly larger covering, e.g., dissociation~\cite{krieger},
chemical design~\cite{laarmann}, and molecular switches~\cite{geppert}.
These research lines within OCT among other applications 
are described in a recent review by Brif {\em et al.}~\cite{octreview2}

On the theoretical side, a single-active-electron approximation is usually 
employed~\cite{decleva} with the full treatment of many-electron effects being 
limited to numerical investigations on very small systems.  
With the aim of studying larger and more complex systems, 
time-dependent density-functional theory~\cite{rungegross,ullrich} (TDDFT) has 
emerged as a computationally efficient method that can, in principle, exactly 
deal with the dynamics of the full many-electron system. 
Within TDDFT the exact time-dependent 
density is obtained from a fictitious system of noninteracting 
electrons moving in an effective time-dependent Kohn-Sham (KS) potential. The KS potential is
a unique functional of the density which, in practice, must be approximated. 
Over the past years different approximation schemes have been developed
and tested showing both promise and future challenges~\cite{mundt,kummel1,helbig1,helbig2,kummel}. 
In addition, very recently, the {\em inverse} problem, i.e., finding the field 
that drives the many-particle system to a desired outcome, has been 
formally solved within a combination of OCT and TDDFT~\cite{tddftoct}.

In the present work we take steps in the practical validation of 
this combination by testing how fields 
optimized in the TDDFT framework (with different functionals) 
perform when applied in the numerically {\em exact} time-dependent Schr\"odinger
equation. This question is of particular relevance from the experimental
point of view; namely, when considering a many-electron system 
beyond the capabilities of a numerically exact treatment, can we 
design usable laser pulses with TDDFT+OCT to be used in an experiment
for an enhanced effect? Our response will be affirmative, but with
important reservations as will be discussed.

We focus on maximizing the total ionization of a model H$_2$-molecule. 
First, a target functional in terms of the density is formulated and 
carefully validated. As our main result we show that the application of 
OCT in conjunction with TDDFT produces pulses that -- when used in the exact system --
lead to considerably higher ionization yields than non-optimized 
single-frequency fields with the same fluence. However, the lack
of correlation effects beyond the simple approximations 
in the used adiabatic functionals is shown
to affect the yield in a negative way. This effect is further exemplified
by calculating the ionization yield of H$_2$ as a function of the (single) 
photon frequency.

The paper is organized as follows. In Sec.~\ref{system} we introduce
the model H$_2$-system as well as the OCT scheme. In Sec.~\ref{results} 
we present our main results for the optimization of the ionization yield
with different XC functionals in TDDFT and compare with the numerically 
exact scheme. We also investigate the single-frequency pulses which further 
underline the importance of correlation effects. The paper is summarized in 
Sec.~\ref{summary}.

\section{System and methodology}\label{system}

\subsection{Model Hamiltonian}\label{model}

The H$_2$-molecule is modeled in terms of a one-dimensional (1D) 
system with a soft-Coulomb interaction between the 
electrons~\cite{jes88,alberto}. This model 
has been shown to capture many qualitative features of the true 3D 
molecule, which is enough also for our purposes. The time-independent 
Hamiltonian of this system is given in Hartree atomic units (a.u.) by 
\be
H=\sum_{i=1}^2\left\{-\frac{1}{2}\frac{\partial^2}{\partial x_i^2}+V_{\rm ext}(x_i)\right\}+V_{ee}(x_1,x_2),
\ee
where the nuclear potential is 
\be
V_{\rm ext}(x_i)=-\frac{1}{\sqrt{(x_i-R/2)^2+a}}-\frac{1}{\sqrt{(x_i+R/2)^2+a}},
\ee
with $x_i$ being the position coordinate of electron $i$, and $a$ a
``softening'' parameter for the electron-nucleus interaction; 
in this work $a=0.9$. Due to the short duration of
the laser pulse (a few fs) the nuclei are considered to be fixed at 
their equilibrium separation of $R=1.5$ a.u. 
The electron-electron interaction is given by a soft-Coulomb interaction 
according to 
\be
V_{ee}(x_i,x_j)=\frac{1}{\sqrt{(x_i-x_j)^2+1}},
\ee
so that here the softening parameter is one. The exact eigenstates 
and eigenenergies can be found by the exact diagonalization of $H$. 
Applying a time-dependent field implies solving the full time-dependent 
Schr\"odinger equation. In this work the system is assumed to be in the 
ground state $\Psi(t=0)=\Phi_0$ when the laser is switched on. 
It is easy to see that, numerically, solving a 1D two-electron problem 
is equivalent to solving a one-electron problem in 2D. Such calculations 
can be carried out using the {\tt OCTOPUS} code~\cite{octopus}, which
is our choice for all the results presented in this work.

\subsection{Kohn-Sham system}

The TDDFT description of the same system uses the existence of an 
independent-particle system evolving according to the KS Hamiltonian, 
\be
H^{\rm KS}=\sum_{i=1}^2\left\{-\frac{1}{2}\frac{\partial^2}{\partial x_i^2}+V_{\rm KS}[n](x_i)\right\},
\ee
that exactly reproduces the true interacting density 
\be
n(x,t)=2\int dx'|\Psi(x,x',t)|^2
\ee
Due to the Runge-Gross theorem~\cite{rungegross} the density uniquely determines all
observables as a function of time. The KS potential 
$V_{\rm KS}$ is normally split into the 
external $V_{\rm ext}$, the Hartree $V_{\rm H}[n]$ and the exchange-correlation 
(XC) potential $V_\xc[n]$. If the external potential is time-dependent, also 
$V_{\rm H}$ and $V_\xc$ become time-dependent. In this work we have tested two 
different approximations to $V_\xc$: the TD exact-exchange (TDEXX) 
approximation and the TD local-density approximation (TDLDA). Both of these 
approximations are adiabatic, i.e., they depend only on the instantaneous 
density (not on its history). For two electrons the TDEXX approximation is 
equivalent to the time-dependent Hartree-Fock approximation of 
many-body perturbation theory and equals 
$V_\x(x,t)=-1/2V_{\rm H}(x,t)=-1/2\int dx' V_{ee}(x,x')n(x',t)$. 
A 1D version of the LDA for soft-Coulomb interactions 
has recently been developed by Helbig {\em et al}.~\cite{helbig1}.
 
The equilibrium properties calculated exactly as well as within 
TDEXX and TDLDA are summarized in Table~\ref{table}. The ionization 
energy $I_p$ is obtained from the highest occupied molecular 
orbital (HOMO) of the unperturbed KS system. We find that TDEXX is 
in good agreement with the exact result. Excitation energies can be calculated 
by means of linear response TDDFT~\cite{pgg96} where only the first functional 
derivative $F_\xc=\d V_\xc/\d n$, the so-called XC kernel, enters. In TDEXX 
$F_\x=-1/2V_{ee}(x,x')$ and we find that the two first excitation energies 
$\D E_1$ and $\D E_1$ agree well with the exact values. 
The ``bare'' KS excitation energies are also presented. TDLDA gives a rather poor 
ionization energy due to the exponential decay of the LDA-XC potential. The same property 
leads to only one empty bound state, which, on the other hand, is rather well described.  

\begin{table}
\caption{Ionization potentials from the HOMO eigenvalue and 
excitation energies in TDEXX and TDLDA as compared to the exact 
results (in eV). 
Also the 'bare' KS-EXX results are presented.}
\begin{center}
\begin{tabular}{c|c|c|c|c}\hline\hline
&Exact & TDEXX & KS-EXX & TDLDA \\\hline
I$_p$&19.32&19.05&19.05&12.52\\
$\D E_1$&12.25&12.52&10.34&11.70\\
$\D E_2$&15.24&15.51&14.97&-\\\hline\hline
\end{tabular}
\end{center}
\label{table}
\end{table}
Even if TDEXX and TDLDA give a reasonable account of the ground -and 
low-lying excited states of the system, it remains to be tested to what extent
non-linear responses can be captured. Previous studies show that, e.g., the 
adiabatic approximation tend to detune resonances in the non-linear regime~\cite{helbig1}.
In the following we will investigate whether it is possible to 
find a common femtosecond laser pulse that can enhance the ionization yield in 
{\em both} the KS system and in the exact system.
\begin{figure}
\includegraphics[width=1.0\columnwidth]{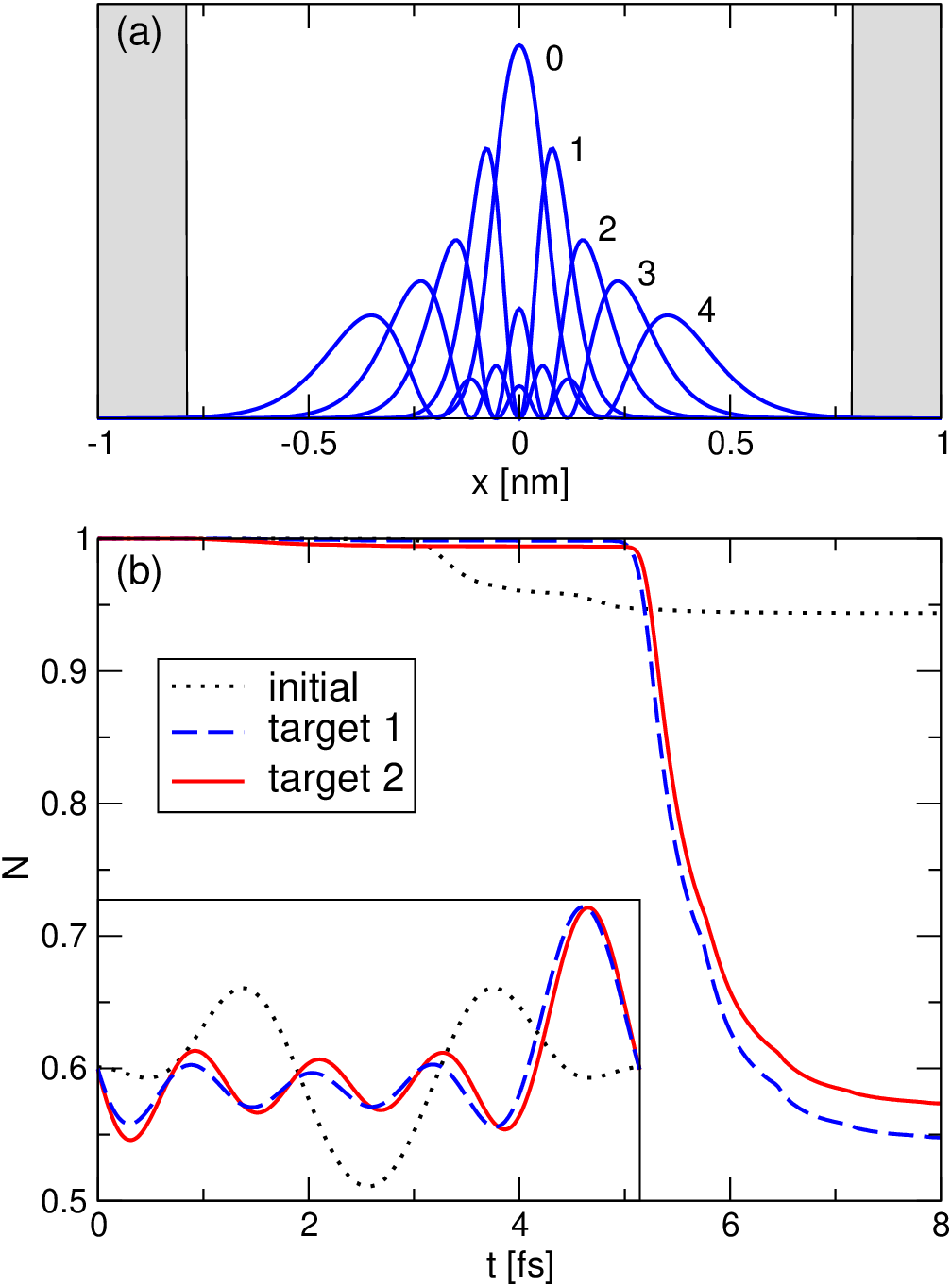}
\caption{(a) Probability densities of the five lowest 
eigenstates in a 1D $H_2^{+}$ molecule. The first target
for ionization is to exclude the occupation of these states
(``target 1''), whereas the alternative target is to
maximize the overlap between the density and the shaded
region (``target 2''). (b) Electron number in $|x|\leq R$ as
a function of time in the presence of the initial pulse (dotted line)
and pulses optimized by using target 1 (dashed line) and target 2
(solid line). The pulse profiles are shown in the inset.}
\label{fig1}
\end{figure}
\subsection{Pulse optimization and the target functional}\label{target}
The laser pulse for the optimization is expressed as
\be
\ve(t)=\sum_{n=1}^M\left[f_n\sqrt{\frac{2}{T}}\cos(\w_nt)+g_n\sqrt{\frac{2}{T}}\sin(\w_nt)\right],
\ee
where the amplitudes $\{f_n,g_n\}$ are varied during the optimization. 
The number of allowed frequencies $M$ is determined by the cut-off 
frequency $\w_{\rm max}$ and the pulse duration $T$, fixed to values
$9.25$ eV and $5.3$ fs, respectively. This gives $M=12$.
The cut-off frequency is thus chosen to be smaller than two times the 
ionization energy and smaller than the first excitation energy. As we will 
see this choice still allows for ionization via excited states. 
The amplitudes are constrained by keeping the fluence, i.e.,
the time-integrated intensity of the laser pulse fixed.

We apply OCT in a so-called direct-optimization scheme presented in
detail in Ref.~\cite{alberto}. In practice, we maximize a merit function
for a set of parameters of the laser pulse by using the
derivative-free NEWUOA algorithm~\cite{newuoa} between  
consecutive time-propagations. Expressing the pulse in a proper 
Fourier basis~\cite{alberto} guarantees that the conditions 
$\int_0^T dt\,\ve(t)=0$ and $\epsilon(0)=\epsilon(T)=0$ are
satisfied.

In order to maximize the ionization yield we need to formulate a target 
functional to be used in the OCT calculation. 
In Ref.~\cite{alberto} the ionization target was expressed as an
{\em exclusion} of a set of lowest bound states.
Here, in order to apply TDDFT, we need to write the target in terms of the
density only. This gives us two choices: (i) we can minimize
the density inside radius $R$ or (ii) we can maximize it
outside $R$ at the end of the pulse. 
In case (i) we minimize the overlap between the density
and a Heaviside step function of the form 
$-\Theta(R-|x|)$, whereas in case (ii) we maximize
the overlap between the density and $\Theta(|x|-R)$.
In principle, these cases are identical, but due to a
finite simulation box and absorbing boundaries we resort
to choice (i) -- apart from the test case described below in this section.
We set $R$ to be equal to the box radius (40 a.u.). 

The ionization probability $P$ can be determined from the remaining 
density in the system in the long-time limit, i.e.,
\be
P=1-\frac{1}{2} \int_{-R}^R dx\,n(x,t\rightarrow\infty). 
\label{ionprob}
\ee
In practice we calculate $P$ at $T=8$ fs, when the density has almost 
fully converged. As the pulse length is fixed to $5.3$ fs, we thus
continue the time-propagation after the field has been switched off,
during which the density continues to evolve in the system.

In Fig.~\ref{fig1} we assess the validity of the density target described 
above by considering the ionization process of a 1D {\em single-electron} 
$H_2^{+}$ molecule. The peak intensity of the initial pulse is 
set to $10^{15}$ W/cm$^2$.
Figure~\ref{fig1}(a) shows the probability densities of the five lowest 
eigenstates. The first target operator (``target 1'') is defined as
$1-\sum_{i=0}^{4}|\Phi_i\big>\big<\Phi_i|$,
i.e., we are attempting to avoid the occupation of the five lowest states.
This type of a target for ionization has been validated in 
previous studies~\cite{alberto}.
Alternatively, ``target 2'' is defined from the density by maximizing
the overlap with the shaded region in Fig.~\ref{fig1}(a) corresponding
to $\Theta(|x|-R)$ [case (ii) above].
To enable a direct comparison between the two targets in this example, 
we use here $R=15$ that approximately agrees with the spatial extent of the 
four lowest eigenstates [Fig.~\ref{fig1}(a)].
\begin{figure*}[t]
\includegraphics[width=0.97\columnwidth, clip=true]{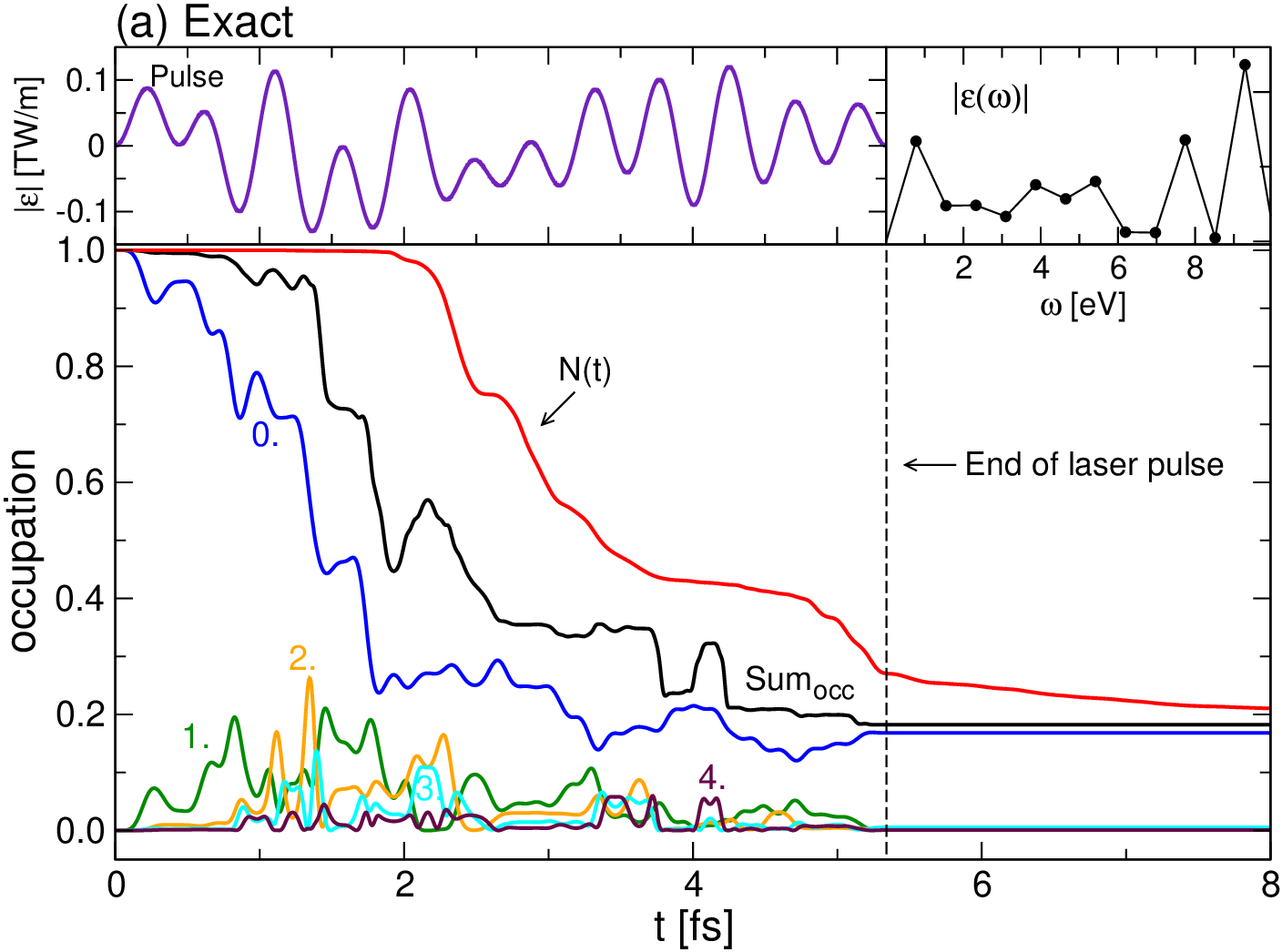}\hspace{10mm}
\includegraphics[width=0.97\columnwidth, clip=true]{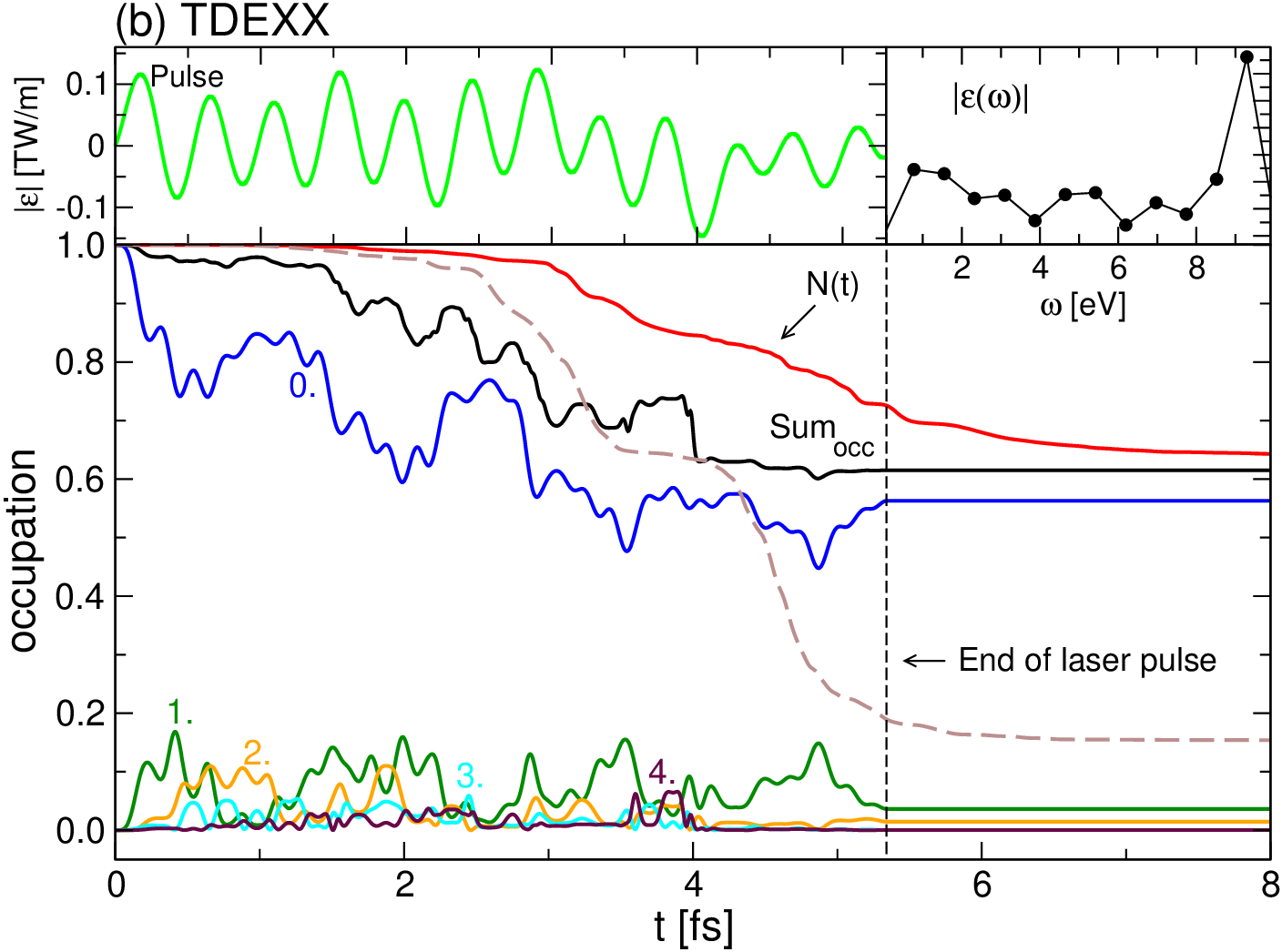}\\\vspace{5mm}
\includegraphics[width=0.97\columnwidth, clip=true]{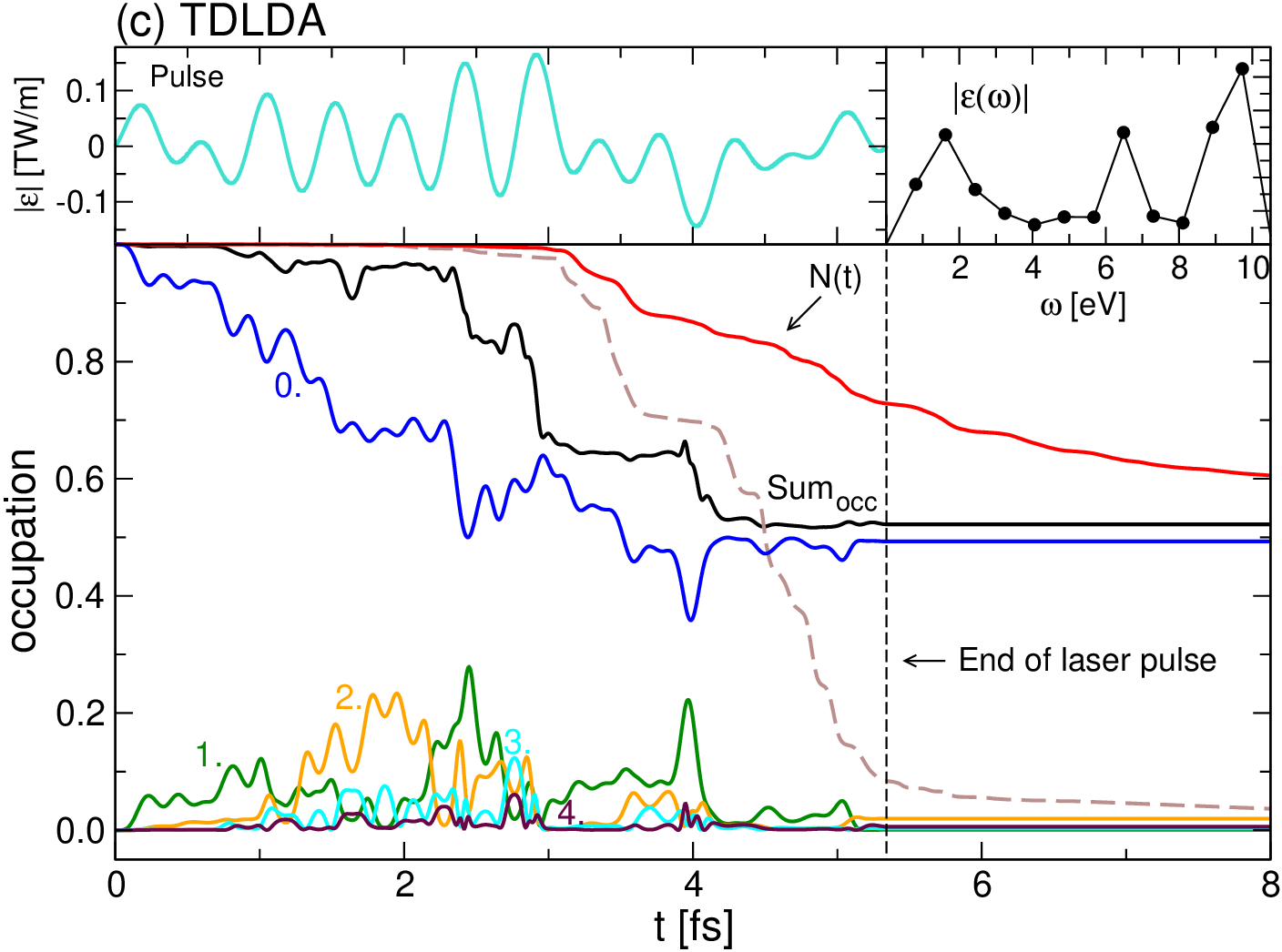}\hspace{10mm}
\includegraphics[width=0.97\columnwidth, clip=true]{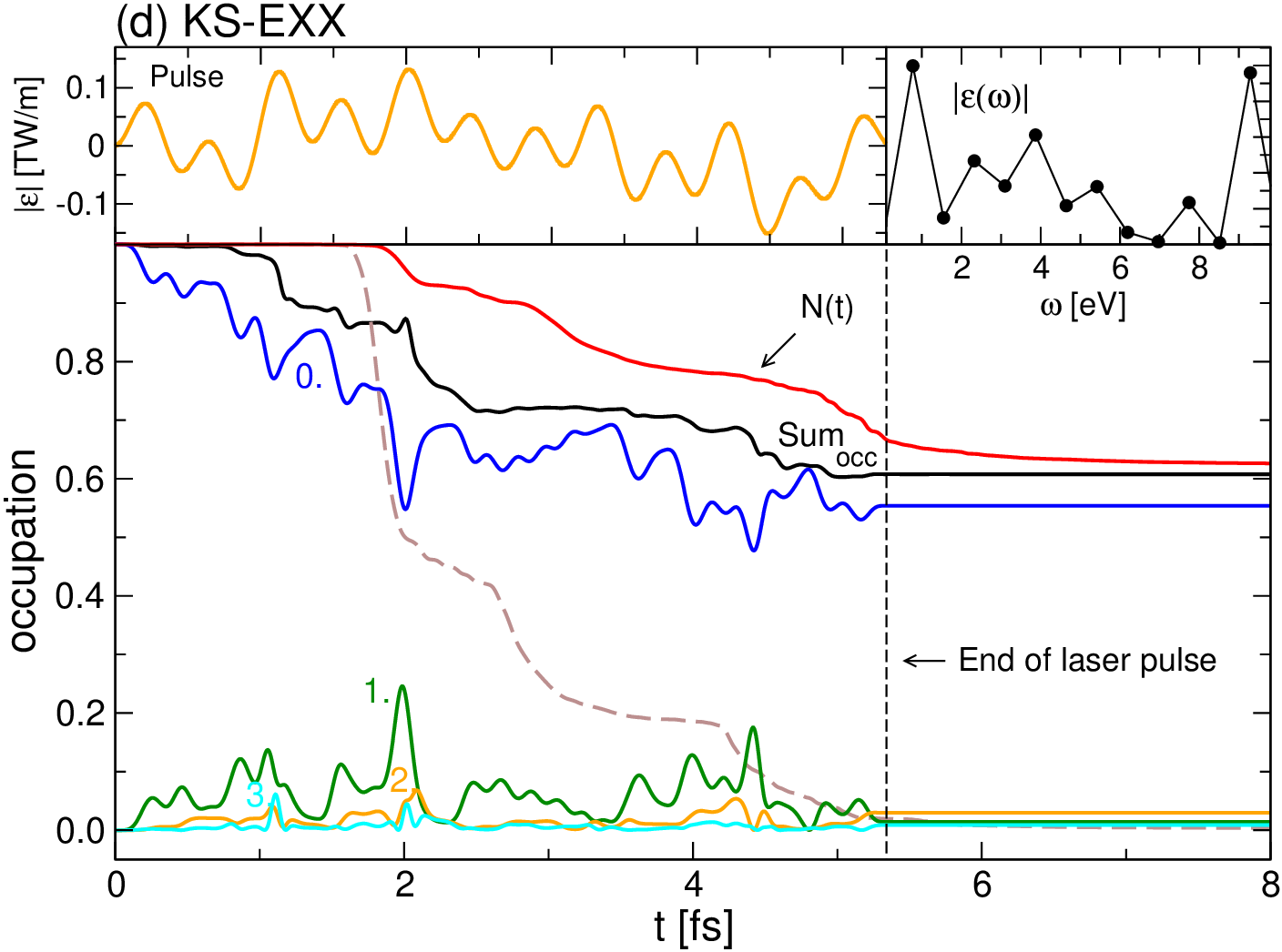}
\caption{Optimized laser pulses (upper left panel), 
their Fourier transforms (upper right panel), and the projections on the excited states as well as the number of particles as a function of time in the exact system (lower panel) from (a) exact optimization, (b) TDEXX, (c) TDLDA, and (d) KS-EXX. The dashed fade line in (b)-(d) is the number of particles $N(t)$ in the corresponding KS system under the influence of the same laser pulse.}
\end{figure*}
As shown in Fig.~\ref{fig1}(b), the results for the decaying 
number of electrons $N$ in $|x|\leq R$ as a function of time, 
as well as for the 
optimized pulses (inset), are almost identical for these two targets.
In both cases the ionization probability is significantly increased 
through optimization. We point out that here the high ionization yield
results primarily from the high intensity peak near the end of the
pulse. Detailed discussion on such an OCT process in the tunneling
regime can be found in Refs.~\cite{alberto} and \cite{lars}.

\section{Results}\label{results}

\subsection{Results for ionization}

Now we switch back to the original two-electron H$_2$ system defined in
Sec.~\ref{model}. Figure 2(a) shows the results obtained from the 
optimization in the {\em exact} system. The optimal laser pulse 
and its Fourier transform are shown in the upper left and right panel, 
respectively. In the lower panel we show the evolution of the 
integrated density (normalized to one) in the system, 
$N(t)=(1/2)\int_{-R}^R dx\,n(x,t)$ (red solid line). 
Thus, in the long-time limit the ionization probability
in Eq.~(\ref{ionprob}) can be expressed as
 $P=1-N(t\rightarrow \infty)$. We also plot the 
projections $|\bra\Psi(t)|\Phi_i\ket|^2$ $[i=0,...,4]$, where $\Psi(t)$ 
is the time-evolved wave function and $\Phi_i$ denotes the $i$:th eigenstate 
of the unperturbed system. The sum of the projections, corresponding
to the total occupation of the four lowest eigenstates, is also 
shown. It can be expected that in the $t\rightarrow \infty$ limit
$N(t)$ approaches the sum of the occupations. This is due to the
fact that after the pulse has been switched off, the part of
the electron density that is {\em not} bound in the lowest states 
eventually propagates {\em into the absorbing boundaries}.
This expectation is confirmed by all the results below.

We see in Fig. 2(s) that already at $t=2$ fs the probability 
of an electron being 
ionized is around $50\%$. The ionization process involves 
mainly $\Phi_1$ and $\Phi_2$, starting with an increased occupation of 
$\Phi_1$ as the ground-state is being deoccupied. Around $t=1$ fs 
we find a transfer from $\Phi_1$ to $\Phi_2$, just before 
the laser reaches its peak intensity where ionization is pronounced.
A similar effect is repeated thereafter.
During the remaining pulse duration, a smooth depopulation of the bound 
states takes place, and at the end only $20\%$ of the ground state is
occupied. Around $t=8$ fs, the $N(t)$ curve converges to 
the sum of the projections
and we find the yield to be 80\%. This should be compared with the random 
initial pulse giving a yield of less than 20\%. 

In Fig. 2(b) we show the results obtained by optimizing the laser pulse 
in the KS system within the TDEXX approximation. By applying this laser 
pulse {\em to the exact system} we find the results of the lower panel. The 
dashed fade line in the background is the $N(t)$ curve of the KS density 
under the influence of the same laser pulse, leading to a 
final yield of $80\%$ in the KS system.
When applied to the exact system the same laser pulse gives a 
yield of around $40\%$. This shows that the densities in the exact and in 
the TDEXX systems behave quite differently, and that the pulse optimized
in the TDEXX scheme has only a limited ionization effect on the exact system.
Despite its limitations, the KS optimization
is seen to produce a better pulse than the random initial guess and as we 
shall see later than pulses containing only a single frequency.
The projections 
on the excited states of the exact system show that also with the TDDFT 
pulse the excited states are involved. The major difference as compared 
to the exact case is that the ground state gets {\em repopulated} after 
being depopulated. The transfer to the second excited state or to the 
continuum is therefore not complete as in the exact case. This 
oscillating behavior is seen throughout the pulse duration and it
appears to prevent complete ionization. 

Figure 2(c) shows the results from the TDLDA optimization. In this case we 
are able find a slightly higher yield of $50\%$ when the pulse is applied 
to the exact system. We also see that 
the oscillating behavior is reduced and the second excited state is better 
exploited than in the TDEXX case. We have also performed an optimization keeping 
the KS-EXX potential fixed in time [Fig. 2(d)]. In that case only the 
first excited state is used. The final yield is, however, 
still as high as $40\%$.

We point out that the obtained pulses are not unique, which is a common
feature of all OCT studies. We can 
therefore not entirely exclude the possibility that there is a pulse 
that can produce the same yield in the approximate as in the 
exact system. The algorithm finds slightly different pulses depending 
on the initial condition for the optimization. 

\subsection{Single-frequency pulses}

In order to gain further insight into the ionization process of H$_2$ 
we test the effect of laser pulses containing only a single 
frequency according to 
$\ve(t)=f(t)\cos(\omega t)$, where the pulse envelope is given by
$f(t) = F_0 \cos[\pi/2(t-3T)/T]$. 
The amplitude $F_0$ is chosen to produce a peak intensity
of $5\times 10^{14}$ W/cm$^2$, and the pulse length is $T=2.65$ fs. 
The total propagation time is 8 fs, so that $f(t) = 0$ at
$t>T$. The fluence is equal to OCT processes described above.
The whole frequency range below the cut-off frequency chosen for the 
optimized laser pulses is scanned, and the ionization yield 
is determined by integrating the density in the simulation box 
as explained in Sec.~\ref{target}. 

\begin{figure}
\includegraphics[width=0.97\columnwidth]{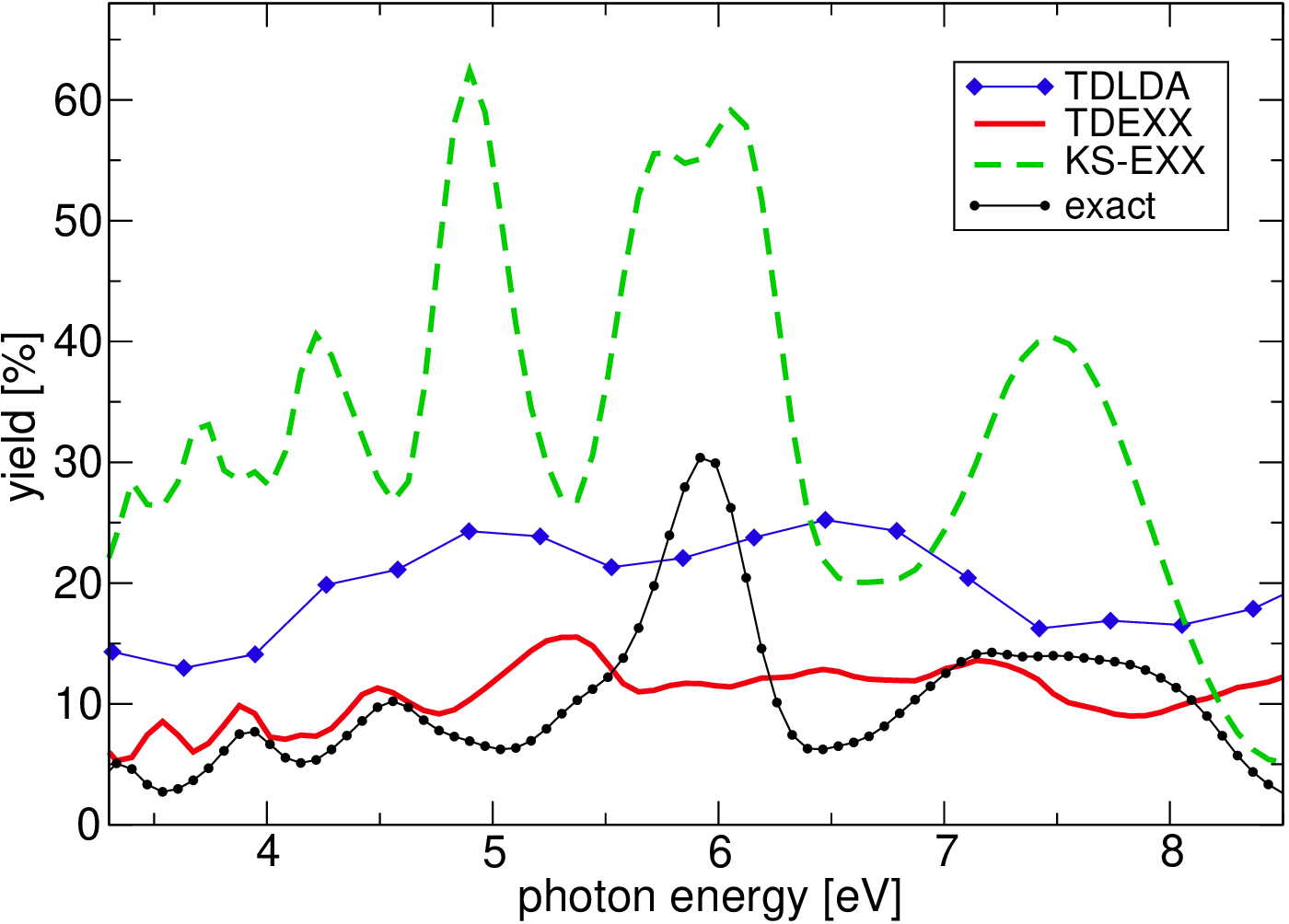}
\caption{Numerically exact result of the total ionization yield 
in a one-dimensional H$_2$ molecule with a soft-Coulomb interaction
as a function of laser frequency (photon energy). The pulse
duration is 2.65 fs and the peak intensity is $5\times 10^{14}$ W/cm$^2$.}
\label{fig3}
\end{figure}

The yield as a function of frequency is plotted in Fig.~\ref{fig3}. 
The exact result 
(black-dotted curve) is compared to the TDEXX result (red solid line), the 
TDLDA (blue dashed line), and the KS-EXX (green dashed line). 
Notice that if the exact time-dependent KS potential were used, the KS 
yield would be equal to the exact yield since only the density is needed 
to determine the ionization yield. Distinct peaks are found at 
certain frequencies. They are strongly emphasized in the noninteracting 
KS-EXX spectrum, which can be considered as our zeroth-order approximation 
with respect to the Coulomb interaction. This 
choice of a zeroth-order system leads to the correct description of the 
ionization energy of the first electron, but misses the fact that the second 
electron should be more bound and much harder to ionize. The role of the 
time-dependent effective potential is to simulate this effect and reduce 
the yield, here by roughly a factor of four. This effect is rather well 
reproduced by the TDEXX. The peaks are also somewhat shifted in the TDEXX
and hence in better agreement with the exact result, at least at low 
frequencies. 

At around $\omega = 6$ eV, we find a sharp peak in the exact spectrum 
that exhibits an asymmetric Fano-type peak. Its location is very close 
to a half of the first discrete excitation energy at $\omega = 12.25$ eV. 
We note, however, that the other peaks cannot be associated to the 
excitation energies of equilibrium system due to the large amplitude 
of the field.
 
In the TDEXX results we find similarities to the exact result, especially
in the low-energy regime, but no clear signatures of a Fano-type resonance. 
This suggests that correlation effects are 
important to describe this resonance. The TDLDA curve shows a suppression 
of the yield but the effect is less accurate as compared to TDEXX. Also here
the resonance is missing. 

The maximum yield we can obtain using a laser pulse with a single frequency 
and peak intensity of 500 TW/cm$^2$ is around $30\%$. Thus, even if we were 
able to locate the resonance, the yield is still lower than 
obtained in the OCT procedure (within all the tested approximations in TDDFT)
that allows for more frequencies.
This motivates the use of OCT for an enhanced yield 
as opposed to an optimal single-frequency pulse.

\subsection{Summary}\label{summary}

We have applied quantum optimal control theory in 
conjunction with time-dependent
density-functional theory (TDDFT) to examine enhanced ionization of a
model H$_2$ molecule. First we have validated the use of a density-based target
for the maximum ionization in the TDDFT framework. According to our main
results, pulse optimization within the (adiabatic) exact-exchange formalism
and the local-density approximation provide reasonable pulses for 
enhanced ionization: when those pulses
are applied to the exact system the yield is considerably increased
with respect to the initial random guess or with respect to the 
single-frequency result. However, we have found that these functionals 
are unable to capture complicated correlation effects
in the system. The presence of these effects becomes clear in the analysis
of the exact ionization yield as a function of the pulse frequency,
revealing, e.g., a sharp resonance. In conclusion, TDDFT may be used
as the first attempt to optimize strong-field effects in atomic systems,
but further work is needed to construct more accurate functionals 
to account for many-particle phenomena at a deeper level.
\begin{acknowledgments}
This work was supported by the Academy of Finland and 
the European Community's FP7 through the CRONOS project, 
grant agreement no. 280879.
We are grateful to CSC -- the Finnish IT Center for
Science -- for computational resources.
\end{acknowledgments}

\end{document}